\documentclass{article}

\usepackage{arxiv}

\usepackage[utf8]{inputenc} 
\usepackage[T1]{fontenc}    
\usepackage{hyperref}       
\usepackage{url}            
\usepackage{booktabs}       
\usepackage{amsfonts}       
\usepackage{nicefrac}       
\usepackage{microtype}      
\usepackage{lipsum}		
\usepackage{graphicx}
\usepackage[square , sort, numbers]{natbib}
\usepackage{doi}
\usepackage[nolist]{acronym}

\title{OPC UA for IO-Link Wireless in a Cyber Physical Finite Element Sensor Network for Shape Measurement}

\date{November 14, 2024}	

\author{ \href{https://orcid.org/0000-0002-5390-3946}{\includegraphics[scale=0.06]{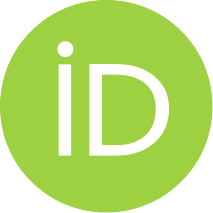}\hspace{1mm}Henry Beuster}\\
	Electrical Measurement Engineering\\
	Helmut-Schmidt-University\\
	Hamburg, Germany\\
	\texttt{henry.beuster@hsu-hh.de} \\
    \And
	\href{https://orcid.org/0009-0006-4224-9906}{\includegraphics[scale=0.06]{orcid.pdf}\hspace{1mm}Lars-Michel Bretthauer}\\
	Electrical Measurement Engineering\\
	Helmut-Schmidt-University\\
	Hamburg, Germany\\
	\texttt{lars.bretthauer@hsu-hh.de} \\
	\And
	{\hspace{1mm}Gerd Scholl}\\
	Electrical Measurement Engineering\\
	Helmut-Schmidt-University\\
	Hamburg, Germany\\
	\texttt{gerd.scholl@hsu-hh.de} \\
}

\renewcommand{\shorttitle}{OPCUA for IOLW in CPFEN}

\hypersetup{
pdftitle={OPC UA for IO-Link Wireless in a Cyber Physical Finite Element Sensor Network for Shape Measurement},
pdfsubject={Sensor Network, IO-Link Wireless, OPC UA, Information Model, Shape Measurement},
pdfauthor={Henry Beuster},
pdfkeywords={Sensor Network, IO-Link Wireless, OPC UA, Information Model, Shape Measurement},
}

\begin{document}
\maketitle

\begin{acronym}
    \acro{iolw}[IOLW]{IO-Link Wireless}
    \acro{plc}[PLC]{programmable logic controller}
    \acro{sfrt}[SFRT]{safety function response time}
    \acro{ism}[ISM]{industrial, scientific and medical}
    \acro{mmtc}[mMTC]{massive machine type communications}
    \acro{iol}[IOL]{IO-Link}
    \acro{lte}[LTE]{long-term evolution}
    \acro{ue}[UE]{user equipment}
    \acro{nsa}[NSA]{non standalone}
    \acro{sa}[SA]{standalone}
    \acro{scs}[SCS]{subcarrier spacing}
    \acro{ofdm}[OFDM]{orthogonal frequency-division multiplexing}
    \acro{urllc}[URLLC]{ultra reliable low latency communications}
    \acro{embb}[eMBB]{enhanced mobile broadband}
    \acro{revpi}[RevPi]{Revolution Pi}
    \acro{vpn}[VPN]{virtual private network}
    \acro{dhcp}[DHCP]{Dynamic Host Configuration Protocol}
    \acro{ip}[IP]{Internet Protocol}
    \acro{rssi}[RSSI]{Received Signal Strength Indicator}
    \acro{iols}[IOLS]{IO-Link Safety}
    \acro{cpfen}[CPFEN]{Cyber Physical Finite Element Sensor Network}
    \acro{wmaster}[W-Master]{Wireless-Master}
\end{acronym}

\begin{abstract}
\footnote{This is the author's version of a paper that has been accepted for Sensor and Measurement Science International 2025 (SMSI2025) and will be presented at the conference in Nuremberg on May 8, 2025.}
This paper presents the integration of OPC UA as a communication protocol in a wireless sensor network and the associated companion specifications as a semantic template for an information model. The \ac{cpfen} for Shape Measurements, a distributed wireless system, uses \acl{iolw} for data transmission at the sensor level, OPC UA provides a unified interface for data access, configuration, monitoring, and calibration tailored to the needs of the \ac{cpfen} for all level above. This opens up additional possibilities, such as integrated quality assurance or creating a digital twin, while improving scalability.
\end{abstract}

\acresetall


\section{Motivation}
\begin{figure}[tb]
    \centering
    \includegraphics[width=0.7\linewidth]{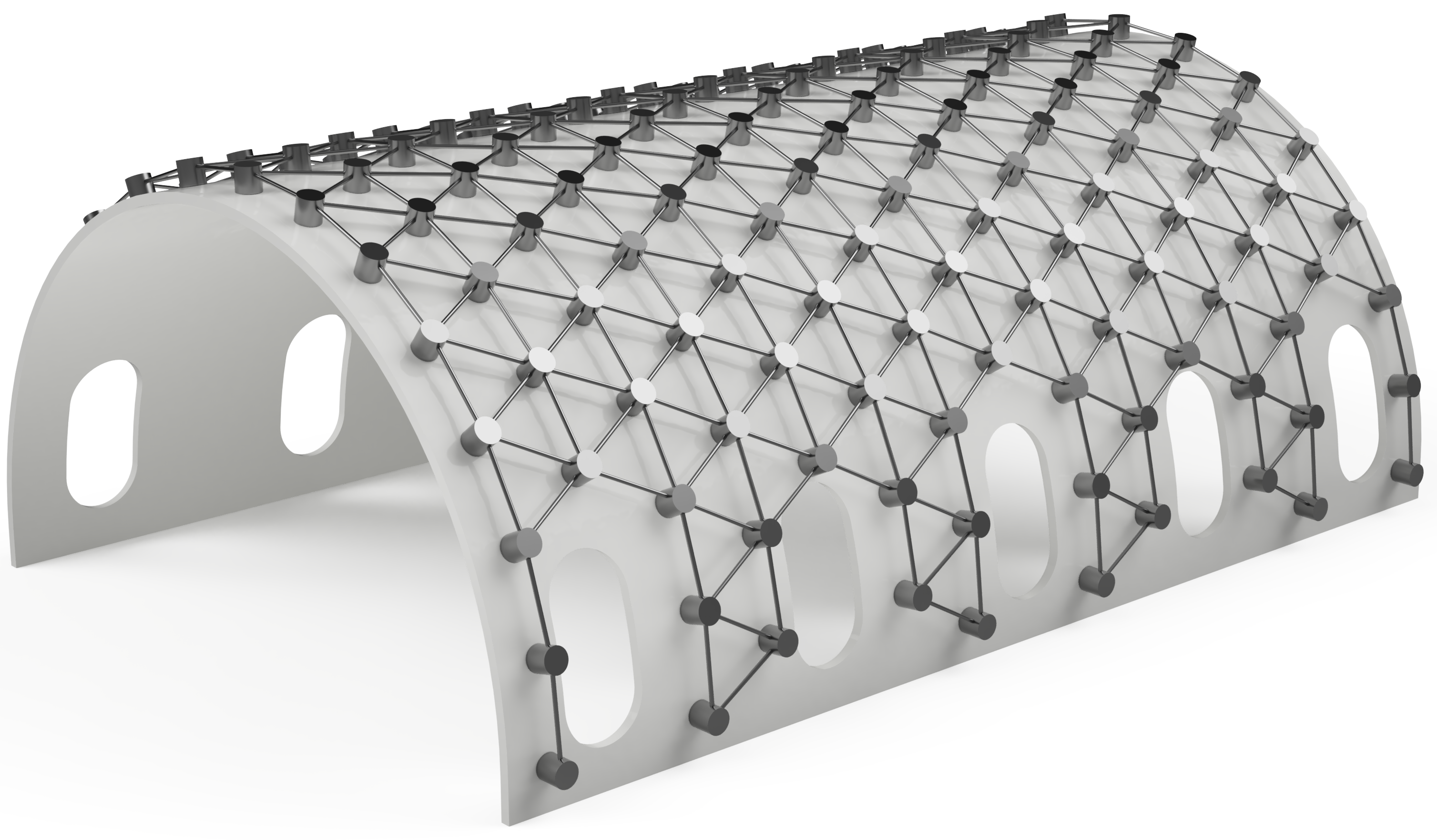}
    \caption{Sensor network attached to the surface of a large structure.}
    \label{fig:sensor}
\end{figure}

The \ac{cpfen} for Shape Measurements proposed in \cite{bretthauer_proposal_2024} is a new approach addressing the challenges of precise real-time shape monitoring in industrial applications. It uses a network of interconnected sensor nodes like a finite element grid attached to the surface, enabling the continuous measurement of shape and deformation in large structures, as depicted in Fig.~\ref{fig:sensor}. This capability is critical for applications such as manufacturing process control and structural health monitoring.
In this network, \ac{iolw} connects sensor nodes to the \ac{wmaster}, enabling the transmission of measurement data without the need for physical cabling. \ac{iolw} also enables centralized handling of calibration, configuration, and extended diagnostic data \cite{beuster_measurements_2024}.
For higher-level communications, OPC UA provides a unified interface as communication framework. It extends across multiple \ac{wmaster}s to the central measurement computer. A customized OPC UA information model is used to meet the specific requirements of the \ac{cpfen}. It provides a semantic structure for organizing sensor data, enabling monitoring, while ensuring easy scalability and remote configurability. This customization enables the integration of applications such as predictive maintenance and process optimization.

\section{Architecture and Sensor Information Model}
\begin{figure}[tb]
    \centering
    \includegraphics[width=0.5\linewidth]{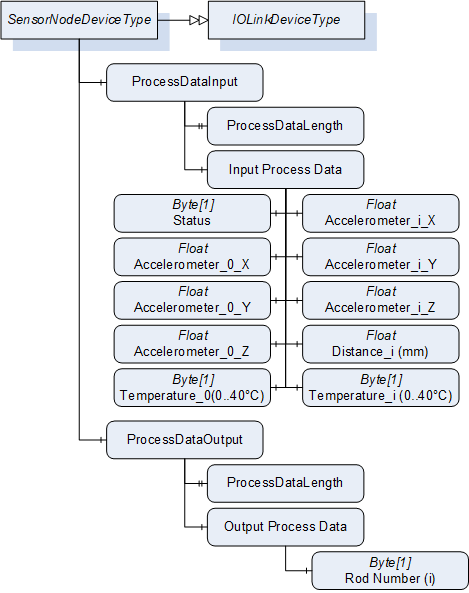}
    \caption{Information model for the SensorNodeDeviceType according to~\cite{noauthor_opc_2018}.}
    \label{fig:model}
\end{figure}

The structure of the \ac{cpfen}, described from the lowest to the highest level, consists of up to 40 sensor nodes per \ac{iolw} \ac{wmaster}. The \ac{wmaster} is implemented on a TI TMDS64EVM evaluation board and provides dedicated cores for real-time processing of \ac{iolw} and sufficient processing power for communication applications such as OPC UA implemented on a non-real-time Linux~\cite{morato_assessment_2021}. In this configuration, the communication application based on the Open62541 stack serves as a gateway, providing both, the OPC UA server functionality and the "Standardized Master Interface" of \ac{iolw} on the same network interface. In the subsequent layer, multiple \ac{wmaster}s can be connected to a control and measurement computer via Ethernet or a corresponding wireless alternative such as 5G~\cite{doebbert_study_2021}. The sensor density is only limited by the capability of an \ac{iolw} cell, defined by the specification~\cite{noauthor_iec_2023}, which allows up to three \ac{wmaster}s with 120 corresponding sensor nodes within a 10\,m cell. 
A customized OPC UA information model, based on the "OPC UA for IO-Link Devices and IO-Link Masters" specification~\cite{noauthor_opc_2018}, is adapted for the sensors of the \ac{cpfen} and the wireless multi-sensor system architecture. It comprises three key elements: firstly, sensor data of each node for easy access and integration; secondly, configuration and calibration parameters for centralized control of settings; and thirdly, diagnostic data for monitoring the sensor health and communication status. The information model provides a flexible and scalable structure that allows efficient expansion and integration with higher-level applications such as quality assurance. Fig.~\ref{fig:model} illustrates the mapping of the process input and output data to the corresponding physical measurement value, which reflects the sensor node structure as a representative of the other aspects mentioned. The sensor nodes consist of a central sensor probe with an acceleration sensor (index 0) and up to six mechanical interconnecting rods, each of which has an acceleration sensor and a distance measurement, of which up to three can be assigned to a node for the data connection (index i 1 to 3).

\section{Conclusion}
The integration of OPC UA with \ac{iolw} in the \ac{cpfen} provides a modern architecture for precise wireless shape monitoring in industrial applications. OPC UA and the information model adapted to the \ac{cpfen} enable standardized communication, configuration and diagnostic handling and facilitate the deployment of the flexible sensor network in large structures.
A promising direction for future research is to improve the real-time capabilities of OPC UA, as investigated by Pfrommer et al.~\cite{pfrommer_open_2018}. They showed that OPC UA PubSub in combination with TSN can meet the demanding real-time requirements in industrial environments, which is also in line with the OPC UA Field eXchange specifications. Time-sensitive applications, such as dynamic condition monitoring of structures, automated quality control, or simply shorter process cycles and therefore also faster production, could be elevated by improved temporal precision and reduced latency. Enriching measurement data with semantics is an enabler for digital twin modeling, rapid integration, and new business models. Rentschler and Drath emphasized the use of AutomationML~\cite{drath_modeling_2018}, while the similar approach with OPC UA also includes the communication protocol, and digital semantic models are largely convertible.

\section*{Funding}
This research is funded by dtec.bw – Digitalization and Technology Research Center of the Bundeswehr. dtec.bw is funded by the European Union – NextGenerationEU (project “Digital Sensor-2-Cloud Campus Platform” (DS2CCP), \href{https://dtecbw.de/home/forschung/hsu/projekt-ds2ccp}{https://dtecbw.de/home/forschung/hsu/projekt-ds2ccp}).

\bibliographystyle{unsrtnat}
\bibliography{references}

\end{document}